\documentclass[fleqn,twoside,twocolumn,nofootinbib,showkeys]{revtex4}
\usepackage[nocpr]{ujp} % \usepackage[cyr]{ujp} for cyrillic
\begin{document}
 \title[New version of $q$-deformed supersymmetric quantum mechanics]%ъюыюэЄшЄєы
{NEW VERSION OF \boldmath$q$-DEFORMED\\ SUPERSYMMETRIC QUANTUM MECHANICS}%
\author{A.M.~Gavrilik}%1 ртЄюЁ
\affiliation{Bogolyubov Institute for Theoretical Physics, Nat. Acad. of Sci. of Ukraine}%шэёЄшЄєЄ
\address{14b, Metrologichna Str., Kyiv 03680, Ukraine}%рфЁхё
\email{omgavr@bitp.kiev.ua}%e-mail
%\affiliation{National University ``L'vivs'ka Politekhnika''}%
%\address{12, Bandera Str., Lviv 79013, Ukraine}%
\author{I.I.~Kachurik}
\affiliation{Bogolyubov Institute for Theoretical Physics, Nat. Acad. of Sci. of Ukraine}%
\address{14b, Metrologichna Str., Kyiv 03680, Ukraine}%
\author{A.V.~Lukash}%
\affiliation{Bogolyubov Institute for Theoretical Physics, Nat. Acad. of Sci. of Ukraine}%
\address{14b, Metrologichna Str., Kyiv 03680, Ukraine}%

\udk{539} %  %72.20.Pa} \razd{\secvii}

\autorcol{A.M.\hspace*{0.7mm}Gavrilik, I.I.\hspace*{0.7mm}Kachurik,
A.V.\hspace*{0.7mm}Lukash}

\setcounter{page}{1025}%

\begin{abstract}
A new version of the $q$-deformed supersymmetric quantum mechanics
($q$-SQM), which is inspired by the Tamm--Dankoff-type (TD-type)
deformation of  quantum harmonic oscillator, is constructed.
 The obtained algebra of $q$-SQM is similar to that in Spiridonov's
 approach. However, within our version of $q$-SQM, the ground state
 found explicitly in the special case of  superpotential yielding
 $q$-superoscillator turns out to be non-Gaussian and takes the
form
 of  special (TD-type) $q$-deformed Gaussian.
\end{abstract}

\vspace{2mm}

\maketitle

\noindent Combining the basic ideas of supersymmetry as incorporated
in supersymmetric quantum mechanics or SQM (here, in one dimension),
on the one hand, and a $q$-deformation that has become very popular
after the discovery of quantum algebras and especially their
Schwinger-type realization through the $q$-deformed oscillator
algebra of Biedenharn and Macfarlane [1, 2], on the other hand, is
important and potentially of much interest.
 Along this root, Spiridonov in Ref.~[3] has proposed some rather general
 deformation of the supersymmetric (SUSY) quantum mechanics [4, 5] on
 the Hilbert space ${\cal H}$ of square integrable functions.
 As a result of the explicit definition of factorization operators realized
 in ${\cal H}$, (at least) two new features appeared. First, the
 familiar SUSY algebra became a $q$-SUSY algebra, i.e., a $q$-deformed extension
 of the SUSY algebra. Second, due to a $q$-deformation of the SUSY
 algebra,
 the conventional degeneracy of the familiar SQM gets lifted.
 Namely, the whole spectrum of $H_+$, the second of superpartner
 Hamiltonians, results from that of the first superpartner Hamiltonian $H_-$
  (save its lowest state) merely by a definite scaling applied to its
  \mbox{spectrum.}\looseness=1

Here, we present a new version of the $q$-deformed supersymmetric
quantum mechanics ($q$-SQM). Its construction is inspired by the
TD-type $q$-deformed oscillator, which was introduced in [6, 7]
and whose unusual properties were studied in [8].
 The scaling operator $T_q$ is an important ingredient of our model.
 It is worth to note that $T_q$ appeared in Spiridonov's version
 of $q$-SQM in such special way that it drops from the bilinears $A A^{\dagger}$
 and $A^{\dagger} A$ of raising/lowering operators.
   Unlike, in our version of $q$-SQM, the scaling operator $T_q$ is
   present, besides the $q$-supercharges, also both in $A A^{\dagger}$,\ $A^{\dagger} A$
   and in the $q$-SUSY Hamiltonian.   An important property of our approach is
   that this formulation naturally leads to a non-Gaussian ground state
   when the superpotential is chosen as that corresponding to the
   $q$-superoscillator.

        Similarly to [3], we define the $q$-SUSY algebra and provide its explicit
        realization on the Hilbert space of square integrable functions.
        Is should be noted that, when a $q$-deformation is
        implanted in the SUSY quantum mechanics, there is no
        degeneracy (natural in standard SUSY models) anymore,
        both in  our version and in previous Spiridonov's one.
        What concerns the latter one, however, we should stress that
        while one sequence of eigenvalues (corresponding to $H_-$) is in fact
        undeformed and coincides with the case of undeformed superoscillator,
        the second one (corresponding to $H_+$) deforms in such way that all
        its eigenvalues result from respective non-deformed eigenvalues
        of $H_-$ by a uniform $q^2$-scaling.

          The raising and lowering operators in [3] entering the definition of
        supercharges as mentioned therein, to generate the $q$-oscillator algebra
       of Biedenharn and Macfarlane.
         The corresponding operators in our version of $q$-SQM obey a much more involved
       deformed oscillator algebra than that of Biedenharn and Macfarlane
       (and hardly known explicitly before). Moreover, we think the model
       given in [3] implies a more complicated (than BM case) $q$-oscillator algebra
       as well, while of course different from ours.\vspace*{-1mm}

\section{SUSYQM (SQM)       %upersymmetric quantum mechanics (SQM)
         and Spiridonov's\\ version of \boldmath$q$-deformed SQM}

   \subsection{\boldmath$N=2$ supersymmetric\\ quantum mechanics}
    $N=2$ SQM is defined by the superalgebra\vspace*{-2mm}
%1
 \begin{equation}
 \begin{array}{l}
 \{Q,Q^{\dagger}\}=H,\quad Q^2=(Q^{\dagger})^2=0, \\[1mm]
   [H,Q]=[Q^{\dagger},H]=0
 \end{array}
    \end{equation}\vspace*{-3mm}

    \noindent
    with the energy of (nondegenerate in case of exact SUSY) ground state $E_{\rm vac}\geq 0$ and twofold degenerate
    spectrum of excited   states.
       The supercharges are conserved, as implied by their commuting
       with the Hamiltonian.
    Throughout the paper, it is understood that
      $ \hat{p}\equiv P=\frac{1}{\rm i} \frac{\rm d}{{\rm d}x}.$

         Recall that the standard representation of SQM is\vspace*{-2mm}
        \[   Q=
        \begin{pmatrix}
            0 & 0 \\
            A^{\dagger} & 0
        \end{pmatrix}\!\!, \quad
        Q^{\dagger}=
        \begin{pmatrix}
            0 & A \\
            0 & 0
        \end{pmatrix}\!\!, \]\vspace*{-7mm}
        \[  A=\frac{\hat{p}-iW(x)}{\sqrt2}, \quad \left[\hat{x},\hat{p}\right]=i, \]\vspace*{-7mm}
        \[      %\\
        H=
        \begin{pmatrix}
            H_+ & 0\\
            0 & H_-
        \end{pmatrix} =
        \begin{pmatrix}
            AA^{\dagger} & 0\\
            0 & A^{\dagger}A
        \end{pmatrix} = \]\vspace*{-7mm}
        \[
        = \frac12   \left(\hat{p}^2+W^2(x)+W^\prime(x)\sigma_3\right)\!,
        \]\vspace*{-9mm}
        \[
        W^\prime(x)\equiv \frac{\rm d}{{\rm d} x}
                                                  W(x), \quad
        \sigma_3=
        \begin{pmatrix}
            1 & 0\\
            0 & -1
        \end{pmatrix}\!\!.
    \]\vspace*{-4mm}

    The superpartner Hamiltonians $H_\pm$ are isospectral, which follows
     from the intertwining relations\vspace*{-2mm}
     %2
    \begin{equation}
        A^{\dagger}H_+ =H_-A^{\dagger} ,\quad H_+ A = AH_-.
    \end{equation}\vspace*{-5mm}

    \noindent
    The choice $W(x)=x$ corresponds to the harmonic oscillator problem
    with standard bosonic algebra\vspace*{-2mm}
    %3
        %\vspace{-3mm}
    \begin{equation}
    [a,a^{\dagger}]=1,\quad [\hat{N},a^{\dagger}]=a^{\dagger},\quad [\hat{N},a]=-a.
    \end{equation}\vspace*{-9mm}

    \subsection{Properties of \boldmath$q$-scaling operator}

 Here and in our main exposition below, we will use like in [3] an important tool of deformed
 SQM,
namely the $q$-scaling operator $T_q$.
    Though its presence in our resulting formulas will be somewhat unconventional,
    this will cause no problems since its action on functions is well defined.

 So, the $q$-scaling operator $T_q$ is defined on smooth functions
 as\vspace*{-2mm}
 %4
    \begin{equation}
        T_q f(x)=f(qx),
    \end{equation}\vspace*{-5mm}

\noindent where $q\in\mathbb{R}$, and $q\geq1$ or $0<q\leq1$.

    The list of its main properties reads
    %5
         \[ T_q F(x)\cdot=\left[T_q F(x)\right]T_q\cdot,\quad
        T_q \frac{\rm d}{{\rm d} x}=
           q^{-1}\frac{\rm d}{{\rm d} x} T_q,
        \]\vspace*{-7mm}
     \begin{equation}
        T_q T_p=T_{qp},\quad T^{-1}_q=T_{q^{-1}},\quad T_1=1,
     \end{equation}\vspace*{-7mm}
       \[
        T_q^{\dagger}=q^{-1} T_q^{-1}, \quad
        (T_q^{\dagger})^{\dagger}=T_q.
          \]
          Note that the operator $\sqrt{q}\, T_q$ is unitary, and the
          operator $\sqrt{q}\, T_q +\frac{1}{\sqrt{q}}T_{q^{-1}}$
          is
          Hermitian.

        The explicit realization of $T_q$ as a pseudo-dif\-fe\-ren\-tial operator
        is\vspace*{-2mm}
        %6
 \begin{equation}
 T_q=\bigl(e^{\ln\! q}\bigr)^{x\frac{\rm d}{{\rm d} x}}=q^{x\frac{\rm d}{{\rm d} x}}.
 \end{equation}
   Obviously, $T_q x = q x$ and $T_q x^m = q^m x^m$ for any integer $m$.

    \subsection{Spiridonov's \boldmath$q$-deformation of SQM:\\ defining relations}

    This deformation is realized by inserting $T_q$ after the factorization operator $A$
    %7
    \begin{equation}
    \begin{array}{l}
    \displaystyle  A\mapsto A_q = \frac1{\sqrt2}\left(\hat{p}-iW(x)\right)\,T_q,
    \\[3mm]
   \displaystyle  A^{\dagger}\mapsto A_q^{\dagger} =
        \frac{q^{-1}}{\sqrt2} T_q^{-1} \left(\hat{p}+iW(x)\right)\!.
    \end{array}
    \end{equation}
    From their products (index $q$ is dropped here and below)
    %8-9
    \begin{align}
        A A^{\dagger} =&\frac{q^{-1}}{2}
        \left(\hat{p}^2+W^2(x)+W^\prime(x)\right)\!,\\[1mm]
        A^{\dagger} A=&\frac q2\left(\hat{p}^2+q^{-2}W^2(q^{-1}x)\right) -
         q^{-1}W^\prime (q^{-1}x)),
    \end{align}
    one can get the $q$-deformed Hamiltonian and supercharges as
    %10
\begin{equation}
\begin{array}{l}
 H=
        \begin{pmatrix}
            H_+ & 0\\
            0 & H_-
        \end{pmatrix}\equiv
        \begin{pmatrix}
            qA A^{\dagger} & 0\\
            0 & q^{-1}A^{\dagger} A
        \end{pmatrix}\!\!, \\[5mm]
   Q=
        \begin{pmatrix}
            0 & 0\\
            A^{\dagger} & 0
        \end{pmatrix}\!\!, \quad
        Q^{\dagger}=
        \begin{pmatrix}
            0 & A\\
            0 & 0
        \end{pmatrix}\!\!. \\
\end{array}
    \end{equation}

      \subsection{Spiridonov's \boldmath$q$-deformation of SQM:\\ algebra of $q$-supersymmetry }

    The above operators satisfy the $q$-deformed $N=2$  SUSY algebra
    %11
    \[  \{Q^{\dagger},Q\}_q= H, \quad \{Q,Q\}_q=\{Q^{\dagger},Q^{\dagger}\}_q=0,
        \]
    \begin{equation}
         [H,Q]_q=[Q^{\dagger}, H]_q=0,
    \end{equation}
    where the commutators and anticommutators are now replaced by the corresponding
    $q$-brackets:
    \[ [X,Y]_q\equiv qXY-q^{-1}YX,\quad [Y,X]_q=-[X,Y]_{q^{-1}},
       \]\vspace*{-7mm}
       \[
        \{X,Y\}_q\equiv qXY+q^{-1}YX,\quad \{Y,X\}_q=\{X,Y\}_{q^{-1}}.
     \]
    As we see, the supercharges are not conserved.

    The intertwining relations for the Hamiltonians $H_\pm$ encoded  in (11)
    obviously change:
    %12
    \begin{equation}
        A^{\dagger} H_+  = q^2 H_- A^{\dagger},\quad  H_+ A = q^2 A H_-.
    \end{equation}
    This implies that $H_-$ and $H_+$ (without the lowest state of $H_-$)
    are not isospectral to each other, but rather $q$-isospectral: the spectrum
    of $H_+$ results from the spectrum of $H_-$ (without its lowest state)
    by applying the uniform $q^2$-scaling, that is,
         \[  H_+\, \psi^{(+)}=E^{(+)}\psi^{(+)}, \quad
        H_-\, \psi^{(-)}=E^{(-)}\psi^{(-)},
    \]\vspace*{-7mm}
    \[ E^{(+)}=q^2\, E^{(-)}, \quad
        \psi^{(+)}\propto A \psi^{(-)}, \quad
        \psi^{(-)}\propto A^{\dagger} \psi^{(+)}.
   \]

%   \frametitle{Spiridonov's $q$-supersymmetric oscillator}
%
%   Possible exception concerns only the lowest level in the same spirit
%   as it was in the undeformed SUSY quantum mechanics. If $A^{\dagger}, A$
%   do not have zero modes then there is one-to-one correspondence between
%   the spectra. This is the case of a spontaneously broken $q$-SUSY
%   because for it $E_{\rm vac}>0$. If $A$ (or, $A^{\dagger}$) has zero mode
%   then $q$-SUSY is exact, $E_{\rm vac}=0$, and $H_-$ (or, $H_+$) has one level
%   less than its superpartner $H_+$ (or, $H_-$).}

   \noindent\underline{\it Special case of $q$-supersymmetric oscillator}

     \vspace{3mm}
    \noindent Consider the simplest physical example of a $q$-su\-per\-os\-ci\-lla\-tor,
    for which the superpotential is $W(x)=-x$ (or $W(x)=x$).
    In that case, we have
    \[   A A^\dagger = \frac{q^{-1}}{2} (\hat{p}^2 + x^2 -1) , \quad
     A^\dagger A = \frac{q}{2} (\hat{p}^2 + q^{-4} x^2 + q^{-2}), \]
    that yields the anticommutator and the commutator
    \[ \{A, A^\dagger\} = \frac{q^{-1}}{2} \Bigl(\!(1+q^2)\hat{p}^2 + (1 +
    q^{-2})x^2\!\Bigr),
    \]\vspace{-7mm}
   \[ [A, A^\dagger] =  \frac{q^{-1}}{2} \Bigl(\!(1-q^2)\hat{p}^2 +
    (1 -  q^{-2})x^2 -2 \Bigr),
    \]
along with such versions of $q$-commutators:
\[ A A^\dagger - q A^\dagger A = \frac12 \Bigl(\!(q^{-1}-q^2)\hat{p}^2 +
    (q^{-1} - q^{-2})x^2 -q^{-1}-1 \!\Bigr) ,
        \]\vspace{-7mm}
\[ q A A^\dagger - A^\dagger A = \frac12 \Bigl(\!(1-q)\hat{p}^2 +
    (1 - q^{-3})x^2 -q^{-1}-1 \!\Bigr) ,
        \]\vspace{-7mm}
\[ q A A^\dagger - q^{-1} A^\dagger A = \frac{1+q^{-2}}{2}
\Bigl(\!(1-q^{-2})x^2 -1 \!\Bigr) ,
        \]\vspace{-7mm}
\[ q^{-1} A A^\dagger - q A^\dagger A = \frac12 \Bigl(\!(q^{-2}-q^2)\hat{p}^2
   - q^{-2} - 1 \!\Bigr) .
        \]
    The Hamiltonian of this $q$-superoscillator takes the form
    (with $I_2$ being a 2$\times$2 unit matrix)
    %13
    \[ 4H=\left[2\hat{p}^2+\left(1+q^{-4}\right)x^2 + 1 - q^{-2} \right]I_2+
    \]\vspace*{-7mm}
 \begin{equation}
        + \left[\left(1-q^{-4}\right)x^2+1+q^{-2}\right]\sigma_3.
    \end{equation}
    It describes a spin-1/2 particle in the harmonic potential
    and with transverse magnetic field.

    {\bf Remark.} It is important to note that the spectrum
    (it is obviously equidistant) of the superpartner Hamiltonian $H^-$
    coincides, up to an overall multiplier $q^{-1}$, with that of the
    corresponding superpartner Hamiltonian in the usual (non-deformed)
    SUSY quantum mechanics (see, e.g., [5]).

    Let us comment on the physical meaning of the deformation
    parameter $q$: it plays a role of some additional interaction constant.
    Note also that this model possesses the exact $q$-deformed SUSY.
    As mentioned in [3], for the value of $q^2$ being a simple rational number,
    the spectrum of the $q$-superoscillator shows an accidental degeneracy.

\section{New version of \boldmath$q$-SQM inspired\\ by the Tamm--Dankoff deformation}

    \subsection{Tamm--Dankoff (TD) deformed oscillator}

    The ``Tamm--Dankoff cutoff'' deformed oscillator is given in
    terms of a $q$-bracket of the TD type:
    \[
        \hat{N} \mapsto \{\hat{N}\}_q \equiv \hat{N} q^{\hat{N}-1},
        \quad  \{\hat{N}\}_q \stackrel{q\to 1}{\longrightarrow} N ,
        \]\vspace*{-7mm}
       \[
        a^{\dagger}a=\{\hat{N}\}_q, \quad a a^{\dagger}=\{\hat{N}+1\}_q,
    \]
    with the algebra (denote $a\equiv a^-$ and $a^{\dagger}\equiv a^+$)
    %14
    \begin{equation}
        a a^{\dagger}-q a^{\dagger}a=q^{\hat{N}},\quad
        \left[ \hat{N}, a^{\pm} \right] = \pm a^{\pm},
        %\quad \left[ \hat{N}, a^{\dagger} \right] = a^{\dagger},
   \end{equation}
 $\hat{N}$ being the number operator,  and the Hamiltonian
\[  H_{\rm oscil.} = \frac12 \bigl( \{N\}_q + \{N+1\}_q \bigr).
\]
   The energy spectrum of this TD-type $q$-deformed os\-cillator,
 \[ E_{n} = \frac12 \bigl(n q^{n-1} + (n+1) q^n \bigr),
 \]
is very special. As  noticed in [8], the TD-type $q$-deformed
oscillator shows various patterns of the accidental pairwise energy
level degeneracy, always within a definite single pair of levels.

    Below, we will study the $q$-deformation of SUSY QM in the spirit
    of the TD-type $q$-deformation,
    %15
    \begin{equation}
        \hat{N} \mapsto \hat{N} q^{\hat{N}-1}  \quad
           \Longleftrightarrow\quad \hat{p} \mapsto q^{-1} \hat{p}\, T_q,
    \end{equation}
    that is, we will merely adopt the replacement of the momentum
    operator just as it is indicated here.%\vspace*{2mm}

    \subsection{Tamm--Dankoff type \boldmath$q$-deformed SQM}
    We start by introducing a $q$-deformation in the momentum part of
    undeformed factorization operators $A$ and $A^\dagger$:
    %16
 \begin{equation}
\begin{array}{l}
 \displaystyle A \mapsto B = \frac{1}{\sqrt 2} \left(T_q\, \hat{p}- iW(x)\right)\!,
 \\[3mm]
 \displaystyle A^{\dagger} \mapsto B^{\dagger} =
        \frac{1}{\sqrt 2} \left( q^{-1} \hat{p} T_q^{-1} + iW(x)\right)\!.
\end{array}
    \end{equation}
    Then,\vspace*{-2mm}
    %17
    \[  BB^{\dagger}=\frac{1}{2q} (q^{-2} \hat{p}^2+qW^2(x)+
     W'(qx)T_q\,+    \]\vspace*{-5mm}
 \begin{equation}
        +\, i W(qx)\hat{p}T_q-i W(x)\hat{p}T_q^{-1} ), %\\
 \end{equation}\vspace*{-5mm}
 %18
  \[  B^{\dagger}B=\frac{1}{2q}( \hat{p}^2+qW^2(x)-
     W'(q^{-1}x)T_q^{-1}\,+   \]\vspace*{-5mm}
      \begin{equation}
      +\,i W(x)\hat{p}T_q-i W(q^{-1}x)\hat{p}T_q^{-1} ).
    \end{equation}
        Setting $B^{\dagger} B\equiv H_-$ and $B B^{\dagger}\equiv H_+$
        we come to the $q$-deformed algebra for $H,Q,Q^\dagger$ (see also Eq. (11)):
   \[  H=
       \begin{pmatrix}
         H_+ & 0\\
         0 & H_-
        \end{pmatrix}\equiv
        \begin{pmatrix}
            qB B^{\dagger} & 0\\
            0 & q^{-1}B^{\dagger} B
        \end{pmatrix}\!\!,
        \]\vspace*{-5mm}
\[  Q=
     \begin{pmatrix}
         0 & 0\\
         B^{\dagger} & 0
     \end{pmatrix}\!\!, \quad
        Q^{\dagger}=
        \begin{pmatrix}
            0 & B\\
            0 & 0
        \end{pmatrix}\!\!;
        \]\vspace*{-5mm}
        \[
        \{Q^{\dagger},Q\}_q= H, \quad \{Q,Q\}_q=\{Q^{\dagger},Q^{\dagger}\}_q=0,\quad
        \]\vspace*{-7mm}
        \[
        [H,Q]_q=[Q^{\dagger}, H]_q=0.
     \]
    As seen, the supercharges in our model are not conserved as well.

    \subsection{Tamm--Dankoff type deformation:
    \boldmath$q$-deformed supersymmetric oscillator}
    We consider a $q$-supersymmetric oscillator with the superpotential
     $W(x) = -x$. In this case,
     %19
    \begin{equation}
        B \!=\! \frac{1}{\sqrt 2} \left(T_q\,\hat{p} -\! i X)\right)\!, ~~
        B^{\dagger} \!=\! \frac{1}{\sqrt 2} \left( q^{-1} \hat{p} T_q^{-1}\! +\! i
        X\right)\!,\!\!\!\!
    \end{equation}
    where $X\equiv x\cdot$.
    Now (16)--(17) turn into
    \[          BB^{\dagger}=
        \frac{1}{2q}\Bigl(\!q^{-2} \hat{p}^2+q X^2
          + qT_q\\
          +i X\hat{p} (q T_q-T_q^{-1})\!\Bigr), \]\vspace*{-7mm}
       \[ B^{\dagger}B=
        \frac{1}{2q}\Bigl(\!\hat{p}^2+qX^2
             -q^{-1}T_q^{-1}\\
             +i X\hat{p} ( T_q-q^{-1}T_q^{-1}) \!\Bigr).
    \]
 From these, different versions of the permutation relation
 involving different ($q$-)commutators  result:
\[ BB^{\dagger}-B^{\dagger}B = \frac{1}{2q}\Bigl(\! (q^{-2}-1) \hat{p}^2 +
 \]\vspace*{-7mm}
 \[ +\, i X\hat{p} (1-q^{-1})  \Bigl.  (q T_q-T_q^{-1})
+ q T_q+q^{-1}T_q^{-1}\!\Bigr), \]\vspace*{-7mm}
   \[ q BB^{\dagger}-q^{-1}B^{\dagger}B
                =\frac{1}{2q}\Bigl(\! (q^{2}-1) X^2
                + \]\vspace*{-7mm}
   \[+ \,i X\hat{p} (q-q^{-2}) \Bigr.
       (q T_q-T_q^{-1})+q^2 T_q+q^{-2}T_q^{-1}\!\Bigr), \]\vspace*{-7mm}
       %%%%%
\[   BB^{\dagger}-q B^{\dagger}B
          =\frac{1}{2q}\Bigl(\!
            (q-q^2) X^2+ \Bigr. \]\vspace*{-7mm}
         \[    +\,  (q^{-2}-q) \hat{p}^2 +
  \Bigl.            + q T_q+T_q^{-1}\!\Bigr) ,       \]\vspace*{-7mm}
    \[   q BB^{\dagger}- B^{\dagger}B
     =\frac{1}{2q}\Bigl(\! (q^{-1}-1)\hat{p}^2 + (q^2-q) X^2+ \Bigr. \]\vspace*{-7mm}
  \[  \Bigl.  +\,{\rm i} X \hat{p}(q-q^{-1})(q T_q - T^{-1}_q) +
  (q^2 T_q+q^{-1}T_q^{-1})\!\Bigr).
    \]

Here we observe the following:

   -- if we use a usual commutator, the dependence on $X^2$ drops on the r.h.s.;

   -- if we use the $q$-commutator $qAB-q^{-1}BA$, the dependence on
   $\hat{p}^2$ drops on the r.h.s.;

   -- if we use the $q$-commutator $AB-qBA$, the terms with ${\rm i}X\hat{p}$
   cancel out.

   \noindent
   As a check of consistency, we verify: in the limit $q\to 1,$ each of
   these
   relations turns into the standard commutation relation $[B , B^{\dagger}]=1$ for the boson
   operators $B$ and $B^{\dagger}$.

\subsection{TD-type deformation:\\ relation with deformed Heisenberg algebra}

We wish to find explicitly the relation of our deformed oscillator
algebra with some version of the deformed Heisenberg algebra, say
along the lines described in [12].
  For this, we solve Eq. (19) for $X$ and $P$ in terms of the
operators $B$,\ $B^\dagger$ (and also $T_q$,\ $T_{q^{-1}}$):
%20
    \[
        P =q\sqrt2\left(\!B\frac{T_q}{1+T_{q^2}}
            + B^{\dagger} \frac{T_q^{-1}}{1+T_{q^{-2}}}\!\right)=
            \]
      \begin{equation}
       =q\sqrt2 (B+B^{\dagger})\frac{1}{T_q+T_{q^{-1}}},
    \end{equation}\vspace*{-7mm}
    %21
   \[
       X={\rm i}\sqrt2 \left(\! B\frac{1}{1+T_{q^2}}
                      - B^{\dagger}\frac{1}{1+T_{q^{-2}}} \!\right)=
                      \]\vspace*{-7mm}
    \begin{equation}
    ={\rm i}\sqrt2 (B T_{q^{-1}} - B^{\dagger}T_q )\,\frac{1}{T_q+T_{q^{-1}}}     .
    \end{equation}
To check once more the Hermiticity of $X$, it is better to use
another formula for $X$ stemming from (19):
%22
   \begin{equation}
  X = \frac{\sqrt2}{2 {\rm i}} (B^{\dagger} -B) + \frac{q^{-1}}{2 {\rm
  i}} P (T_q - T_{q^{-1}}).
   \end{equation}
Now the Hermiticity follows from the skew-Hermiticity of
$B^{\dagger}-B$ and skew-Hermiticity of the product operator $P (T_q
- T_{q^{-1}})$. Likewise, the Hermiticity of $P$ stems, see (20),
from that of
 $B + B^{\dagger}$ and the fact that the product $P(T_q + T_{q^{-1}})$
 is Hermitian.

The operators $P$ and $X$ can be expressed through $q$- or
$q^{-1}$-commutators of the operators $B$ and $T_q^{-1}$ as
follows:
\[  P=\frac{\sqrt2}{q^{-1}-q} [B,T_q^{-1}]_q , \quad
  X= \frac{{\rm i}\sqrt2}{q-q^{-1}} T_q [B,T_q^{-1}]_{q^{-1}} ,
\]
where $[A,B]_q\equiv AB-qBA$.
  With the equality $[A,B]_q A^{-1}=A[B,A^{-1}]_q$ taken into account,
  we have yet another formulas for $X$:
  \[  X= \frac{{\rm i}\sqrt2}{q-q^{-1}} [T_q,B]_{q^{-1}} T_q^{-1}=
  \frac{{\rm i}\sqrt2}{1-q^2} T_q [T_q^{-1},B]_q = \]\vspace*{-7mm}
  \[ =
  \frac{{\rm i}\sqrt2}{1-q^2} [B,T_q]_q T_q^{-1} .
\]
From these expressions after some algebra, we obtain (note that
$[X,P]={\rm i}$) is intact:
   \[    {\rm i} =  X P - P X =
 \frac{2{\rm i}}{(q-q^{-1})^2} \left( T_q^{-1} [T_q,B]_q
 [B,T_q^{-1}]_{q^{-1}}\right. \!- \]\vspace*{-7mm}
 \[ \left. -\, T_q [B,T_q^{-1}]_{q^{-1}}  [B,T_q^{-1}]_q
  \right) \!.
 \]
That implies the validity of two identities:
\[  T_q^{-1} [T_q,B]_q
 [B,T_q^{-1}]_{q^{-1}} - T_q [B,T_q^{-1}]_{q^{-1}}  [B,T_q^{-1}]_q
= \]\vspace*{-7mm}
\[ = (q-q^{-1})^2/2 ,   \]\vspace*{-7mm}
\[  T_q [T_q^{-1},B]_q [B,T_q^{-1}]_q - T_q^{-1} [T_q, B]_q [T_q^{-1},B]_q = \]\vspace*{-7mm}
 \[  = q(q-q^{-1})^2/2 .  \]
On the other hand, we can deduce a $q$-deformed extension of
Heisenberg algebra for the pair of operators $X$ and $\tilde{P} =
T_q P = q^{-1} P T_q$, which is
%23
\begin{equation}
[X,\tilde{P}] =  q^{-1}\bigl({\rm i}T_q+(1-q)\tilde{P}X\bigr) .
\end{equation}
If we compare (23) with the known deformations of the Heisenberg
algebra, e.g., those from [9], we notice the presence of a
$q$-scaling operator $T_q$ times ${\rm i} q^{-1}$ and of the
bilinear $\tilde{P}X$ multiplied by $1-q$.

  Written through the $q$-commutator, it takes another simpler
  though equivalent form
  %24
\begin{equation}
X\tilde{P}-\frac1q\tilde{P}X = \frac{\rm i}{q}T_q .
\end{equation}
  Clearly, in the limit $q\to 1,$ the r.h.s. of both (23) and (24) turns
  into familiar ``${\rm i}$''.

%   \[
%   B = \frac{1}{\sqrt 2} \left(T_q\, p- iW(x)\right), \qquad
%   B^{\dagger} = \frac{1}{\sqrt 2} \left( q^{2} p T_q^{-1} + iW(x)\right).
%   \]

\vspace{1mm}
\subsection{Ground state\\ of TD-type \boldmath$q$-superoscillator}

%%%--------------------------------------------------------------
    Let us find the ground state (zero mode) for the ladder operators $B$ and
    $B^{\dagger}$, namely,
    %25
    \begin{equation}
        \begin{cases}
            B f(x)=0 \vspace{0mm} \\
            B^{\dagger} \tilde{f}(x)\!=\!0
        \end{cases}  \hspace{-3mm} \Rightarrow
        \begin{cases}
            T_q f'(x)+xf(x)=0, \\   %%\quad \forall x\\
            q^{-2}T_{q^{-1}}\tilde{f}'(x)\!-\!x\tilde{f}(x)=0;  %%\quad %\forall x
        \end{cases}
          \end{equation}          \vspace*{-5mm}
          %26
          \begin{equation}
                 \begin{cases}\!
            f(x) = \sum\limits_{k=0}^\infty{C_k x^k} ,\\[3mm]
        \!\tilde{f}(x) = \sum\limits_{k=0}^\infty{\tilde{C}_k x^k} .
        \end{cases}
    \end{equation}
    These turn into  recurrence relations for the expansion coefficients
    %27
    \begin{equation}
        \begin{cases}
           \! f(x):
            \begin{cases}
                C_k+(k+2)q^{k+1}C_{k+2}=0 ,\\
                C_1=0 ,
            \end{cases}\vspace{2mm}\\
           \! \tilde{f}(x):
            \begin{cases}
                -\tilde{C}_k+(k+2)q^{-k-3}\tilde{C}_{k+2}=0 ,\\
                \tilde{C}_1=0 .
            \end{cases}
        \end{cases}
    \end{equation}
   These can be solved (e.g., by using Mathematica), that gives
    %28
    \begin{equation}
        \begin{cases}
f(x)=C_0\sum\limits_{k=0}^\infty{\frac{ q^{-k^{2}}}{k!}\left(-\frac{x^2}{2}\right)^{k}}\\
\tilde{f}(x)=\tilde{C}_0\sum\limits_{k=0}^\infty{\frac{
q^{k(k+2)}}{k!}\left(\frac{x^2}{2}\right)^{k}}
        \end{cases}
    \hspace{-2mm}    \xrightarrow{q\rightarrow1}
        \begin{cases}
            f(x)= e^{-\frac{x^2}{2}},\\
            \tilde{f}(x)= e^{\frac{x^2}{2}}.
        \end{cases}
    \end{equation}
    Here, $C_0$ and $\tilde{C}_0$ are arbitrary constants depending on the
    deformation parameter $q$,  with the obvious property
    $\{C_0,\tilde{C}_0\}\xrightarrow{q\rightarrow1}1$
    in order to recover the undeformed case.
    From these two functions, only $f(x)$ at $q\to1$ recovers the ground state of  standard
    supersymmetric oscillator.
     Moreover, the function $\tilde{f}(x)$ in (28) is not square integrable.
     Thus, we find
     %29
    \begin{equation}
\psi_0=C_0\sum\limits_{k=0}^\infty{\frac{
q^{-k^{2}}}{k!}\left(\!-\frac{x^2}{2}\!\right)^{\!\!k}}\!\! .
    \end{equation}
 We have obtained the unique, i.e. non-degenerate, ground state
 as the (non-Gaussian) eigenfunction of $B^-$ and hence of $H^-$.
   Below, it will be shown that this non-Gaussian wave function
   naturally takes the form of a specially introduced  TD-type $q$-deformed Gaussian exponent.

\subsection{Elements of TD-analysis}

   We will need some more elements of TD-analysis.
  First,  let us introduce the $q$-number of the Tamm--Dankoff type
    %30
    \begin{equation}
        \left( n \right]_q      \equiv     %%\eqdef
        n q^{n-1}.
    \end{equation}
    This form can be easily obtained from the $(p,q)$-number
    $[n]_{p,q}=\frac{p^{n}-q^{n}}{p-q}$ by taking the limit $p\to q$.\\
    The TD-factorial is given as
    %31
    \begin{equation}
        \begin{array}{l}
     (n]_q! = (1]_q (2]_q\,\mbox{...}\,(n]_q = q^{n(n-1)/2} n!, \\[2mm]
          (0]_q!=0!=1, \quad (1]_q!=1!=1.
        \end{array}
    \end{equation}
    Remembering that $q^{z\frac{\rm d}{{\rm d} z}}f(z)=f(qz)$, one can introduce
    the TD-derivative
    %32
    \begin{equation}\label{TDderiv}
    z\mathcal{D}^{\rm (TD)}_z   \equiv
    z\frac{\rm d}{{\rm d} z}q^{z\frac{\rm d}{{\rm d} z}-1},
    \end{equation}
    which acts on monomials as
    %33
    \begin{equation}\label{TDmon}
        \mathcal{D}^{\rm (TD)}_z z^n = (n]_q z^{n-1}.
    \end{equation}
    From these definitions, it is natural to introduce the TD-exponent\vspace*{-2mm}
    \begin{equation}
    %34
        \exp^{\rm (TD)}_q(z)  \equiv
        \sum^{\infty}_{n=0}\frac{z^n}{(n]!} = \sum^{\infty}_{n=0}\frac{q^{-n(n-1)/2}}{n!}z^n,
    \end{equation}
    with the property
    %35
    \begin{equation}
        \mathcal{D}^{\rm (TD)}_z \exp^{\rm (TD)}_q(\alpha z) = \alpha \exp^{\rm (TD)}_q(\alpha z).
    \end{equation}
    In the ${q\!\to\!1}$ (no-deformation) limit:
    $e^{\rm (TD)}_q(z)\!\xrightarrow{q\to1}\!e^{z},$
$    \mathcal{D}^{\rm (TD)}_z\xrightarrow{q\to1}\frac{\rm d}{{\rm d}
z}$.

    The ground state $\psi_0$ has a natural record in terms of the TD-exponent
    %36
    \begin{equation}
        \psi_0(z)=C_0 \exp^{\rm (TD)}_{q^{2}}\left(\!-\frac 12 q^{-1} z^2\!\right)\!.
    \end{equation}
    In the limit ${q\!\to\!1},$ we have $ \psi_0(z)\rightarrow C_0 \exp{(-\frac{z^2}{2})}$.

 \subsection{Relation to bibasic\\ and twin-basic hypergeometric functions}
    First, let us recall the $(p,q)$-exponent
    %37
    \begin{equation}\label{pqexponent}
        \exp_{p,q}(z) = \sum^{\infty}_{n=0}\frac{z^n}{[n]_{p,q}!},
    \end{equation}
    where $[n]_{p,q}! = [1]_{p,q} [2]_{p,q} \mbox{...} [n]_{p,q}$, and the $(p,q)$-number
    was given after Eq. (30).
      The TD-exponent can be recovered from \eqref{pqexponent} in the $p\to q$ limit.
    The family of the so-called twin-basic hypergeometric functions is given as
    %38
    \[     _{r}\Phi_{s}(\{a, b\};\{c, d\};(p,q);z)\equiv   \]\vspace*{-7mm}
    \[
\equiv\sum_{n=0}^{\infty}{\frac{\left((a_1,b_1),(a_2,b_2),\mbox{...},(a_r,b_r);(p,q)\right)_{n}}
 {\left((c_1,d_1),(c_2,d_2),\mbox{...},(c_s,d_s);(p,q)\right)_{n}}}\times
    \]\vspace*{-5mm}
\begin{equation}   \label{twin}
 \times\frac{\left[ (-1)^{n} \left( q/p \right)^{n(n-1)/2}  \right]^{1+s-r}}
 {\left( (p,q);(p,q) \right)_n} z^{n}\!,
 \end{equation}
    where we introduced the shorthand notations
    \begin{gather*}
 \{a,b\} \doteq \left( (a_1,b_1),(a_2,b_2),\mbox{...},(a_r,b_r)\right)  \\
 \left((a_1,b_1),(a_2,b_2),\mbox{...},(a_r,b_r);(p,q)\right)_{n} \doteq \\
\doteq \left( (a_1,b_1);(p,q) \right)_n
\left( (a_2,b_2);(p,q) \right)_n\mbox{...}\\
\mbox{...}\left( (a_r,b_r);(p,q) \right)_n\!,\\
        \left( (a,b);(p,q) \right)_n \doteq (a-b)(ap-bq)(ap^2-bq^2)\mbox{...}\\
      \mbox{...}  (ap^{n-1}-bq^{n-1}),\quad \left( (a,b);(p,q) \right)_0 = 1 .
    \end{gather*}
    Some special cases are:
    %39-44
    \begin{align}
&\left( (0,b);(p,q) \right)_n = (-b)^n q^{n(n-1)/2},\\
&\left( (a,0);(p,q) \right)_n = a^n p^{n(n-1)/2},\\
&\left( (a,b);(q,q) \right)_n = (a-b)^n q^{n(n-1)/2},\\
 &\left( (a,a);(p,q) \right)_n = 0,\\
&\left( (p,q);(p,q) \right)_n = (p-q)^n [n]_{p,q}, \\
&\left( (q^{-1},q);(q^{-1},q) \right)_n = \left(q-q^{-1}\right)^n
[n]_q!.
    \end{align}
    The requirement of convergence of \eqref{twin} implies that $|q/p| < 1$ and also $|z| < 1$.
    Now, we establish the relation between different deformed exponents and the
     twin-basic hypergeometric function $_{1}\Phi_{1}$, namely,
     %45-47
 \begin{gather}
        \exp_{p,q}(z) = {}_{1}\Phi_{1}( (1,0);(0,1);(p,q);(p-q)z
        ),\\[3mm]
        \exp_{q}(z) = {}_{1}\Phi_{1}( (1,0);(0,1);(q^{-1},q);(q^{-1}-q)z
        ),\!\!\!\!\\[3mm]
    \exp_{q}^{\rm (TD)}(z) = \lim_{p\to q}{}_{1}\Phi_{1}( (1,0);(0,1);(p,q);(p-q)z ).
    \end{gather}
    That is, when the limit on the r.h.s. of (47) is performed,
    we obtain nothing but the TD-type $q$-exponent from (36).

    On the other hand, the TD-exponent can be expressed as some
    special
    case of bibasic hypergeometric functions (see, e.g.,
    [10, 11]):
    %48
    \[     _{r,r'}\mathcal{F}_{s,s'}(\underline a,\underline c;\underline b,\underline
 d;(p,q);z)\equiv\]\vspace*{-7mm}   %\\
 \[\equiv\sum_{n=0}^{\infty}{\frac{(a_1,a_2,\mbox{...},a_r;p)_{n}
 (c_1,c_2,\mbox{...},c_{r'};q)_{n}}{(b_1,b_2,\mbox{...},b_s;p)_{n}(d_1,d_2,\mbox{...},d_{s'};q)_{n}\,(q;q)_n}}\times\]
\vspace{-7mm}
\begin{equation}
\times\left[ (-1)^{n} p^{n(n-1)/2}  \right]^{1+s-r}  \left[ (-1)^{n}
q^{n(n-1)/2}  \right]^{s'-r'}\! z^{n}.
    \end{equation}
    Here $\underline a = (a_1,a_2,\mbox{...},a_r)$
    and $(a_1,a_2,\mbox{...},a_r;p)_{n}\equiv$\linebreak
    \mbox{$\equiv\!(a_1;p)_n(a_2;p)_n\,\mbox{...}\,(a_r;p)_n$;}
    and
    $\left(a;q\right)_n\!\equiv\!\frac{(a;q)_{\infty }}{(a q^n;q)_{\infty }}$
    are the $q$-Pochhammer symbols.
 Then we have
    \vspace{-1mm}
    \begin{equation}
        \exp^{\rm (TD)}_{q^{}}(z)\equiv\
        \lim_{p\rightarrow1}\
        _{0,0}\mathcal{F}_{0,1}(-,-;-,0;p,q^{-1};(1-p)z),
    \end{equation}
 where it is taken into account that
 \mbox{$\frac{(p,p)_n}{(1-p )^n}\!\xrightarrow{p\to1}\! n!$}.
Now it is clear that, using (49), our ground-state wave function
$\psi_0$ in (36) can be presented as a particular (limit of) bibasic
hypergeometric function.

\section{Concluding Remarks}

 In this work, we have presented a new version of the $q$-deformed
 supersymmetric quantum mechanics and a new $q$-deformed
 superoscillator.
   Our way of deformation is rooted in the special Tamm--Dankoff-type
  deformation of  quantum harmonic oscillator.
  Though our main defining relation differs from
  Spiridonov's variant of $q$-SQM
  only slightly at first sight (compare (7) and (16)), the consequences are more principal.
     The basic distinction lies in the presence, in our case,
     of the scaling operator $T_q$ (or $T_{q^{-1}}$) in all subsequent formulas:
  for bilinears $B B^\dagger$, $B^\dagger B$, $q$-supercharges, and
  $q$-super-Hamiltonian, while the analogous operators do not involve $T_q$
  in Spiridonov's version (only traces of its action can be seen in these operators).
  The second important distinction concerns the ground states in the two versions:
  it is a usual Gaussian for the superpotential $W(x)=-x$ of a superoscillator
  in Spiridonov's case, and the special TD type of a $q$-Gaussian in
  our paper, see \mbox{(36)--(37) above}.%\looseness=1

At last, let us make three final remarks:

     -- In the version of $q$-SQM presented above, the $q$-SUSY algebra is
     exact,
  as seen from the relations at the end of subsection 2.2, and there is exactly
  one state with the lowest (zero) energy. However, the two-fold degeneracies of exited states of standard
  SUSY models are lifted. Moreover, the whole set of ($q$-dependent)
  eigenvalues of the $q$-superpartner Hamiltonian $H_+$ is obtained
  from the set of $q$-superpartner Hamiltonian $H_-$ eigenvalues (other than zero
  and as well $q$-dependent) merely by the $q^2$-scaling.

   --  The particular $W(x)=\pm x$ case of a $q$-deformed superoscillator
        was considered, and the ground state is found and expressed
      through the $p\to q$ limit of the bibasic $p,q$-hypergeometric function, or
      through the appropriate limit of the twin-basic hypergeometric function.

   -- A complete system of eigenfunctions for the excited states
      of such a $q$-superoscillator yet remains to be found, and it is
      an exciting problem!

   -- A very interesting question is, of course, about a possible
   relation(s), if any, to the recently developed nonlinear extensions of
   Supersymmetric Quantum Mechanics, see, e.g., [9].

\vspace{3mm} {\it We are grateful to participants of QGQIS-2013 for
useful discussion of the results. This work was supported in part by
the Special Program of the Division of Physics and Astronomy of the
NAS of Ukraine.}

%\vspace{-3mm}
%\section*{References}

\vspace*{-5mm}
\rezume{%
О.М. Гаврилик, І.І. Качурик, О.В. Лукаш}{НОВА ВЕРСІЯ
$Q$-ДЕФОРМОВАНОЇ\\ СУПЕРСИМЕТРИЧНОЇ КВАНТОВОЇ МЕХАНІКИ}
{Запропонована нова версія $q$-деформованої суперсиметричної
квантової механіки ($q$-СКМ), інспірована $q$-деформацією квантового
гармонічного осцилятора у формі Тама--Данкова. Отримано алгебру
$q$-СКМ, яка за виглядом подібна до отриманої у підході Спірідонова.
Однак, у рамках нашої версії $q$-СКМ, найнижчий стан для часткового
випадку суперпотенціалу, що відповідає  $q$-суперосцилятору,
знайдено явно~-- він відмінний від гаусіана, і має вигляд
спеціальної (типу Тама--Данкова) $q$-деформації гаусової експоненти.
Встановлено зв'язок останньої з частковими випадками (границями)
бібазисної, а також і твін-базисної узагальнених гіпергеометричних
функцій.}

\end{document}